\documentclass[pra,twocolumn,showpacs,superscriptaddress]{revtex4-1}
\usepackage{makeidx}
\usepackage[utf8]{inputenc}
\usepackage{graphicx}
\usepackage{amsmath}
\usepackage{mathrsfs}
\usepackage{graphicx}
\usepackage{braket}
\usepackage[subrefformat=parens,labelformat=parens,caption=false]{subfig}
\usepackage{dcolumn}
\usepackage{bm}
\usepackage{bbm}
\usepackage{color}
\usepackage[dvipsnames]{xcolor}
\usepackage{esint}
\usepackage{lineno}
\usepackage{verbatim}
\usepackage[breaklinks,
            colorlinks,
            urlcolor=blue,
            linkcolor=blue,
            anchorcolor=blue,
            citecolor=blue]{hyperref}

\DeclareMathOperator{\Tr}{Tr}

\usepackage{soul}

\begin{document}

\title{Long-distance dissipation-assisted transport of  entangled states via a chiral waveguide}

\author{Wai-Keong Mok}
\email{waikeong\_mok@u.nus.edu}
\affiliation{Department of Electronics and Photonics, Institute of High Performance Computing, 1 Fusionopolis Way, 16-16 Connexis,
Singapore 138632, Singapore}
\affiliation{Centre for Quantum Technologies, National University of Singapore, 3 Science Drive 2, Singapore 117543}
\author{Davit Aghamalyan}
\affiliation{Centre for Quantum Technologies, National University of Singapore, 3 Science Drive 2, Singapore 117543}
\author{Jia-Bin You}
\affiliation{Department of Electronics and Photonics, Institute of High Performance Computing, 1 Fusionopolis Way, 16-16 Connexis,
Singapore 138632, Singapore}
\author{Tobias Haug}
\affiliation{Centre for Quantum Technologies, National University of Singapore, 3 Science Drive 2, Singapore 117543}
\author{Wenzu Zhang}
\affiliation{Department of Electronics and Photonics, Institute of High Performance Computing, 1 Fusionopolis Way, 16-16 Connexis,
Singapore 138632, Singapore}
\author{Ching Eng Png}
\affiliation{Department of Electronics and Photonics, Institute of High Performance Computing, 1 Fusionopolis Way, 16-16 Connexis,
Singapore 138632, Singapore}
\author{Leong-Chuan Kwek}
\affiliation{Centre for Quantum Technologies, National University of Singapore, 3 Science Drive 2, Singapore 117543}
\affiliation{MajuLab, CNRS-UNS-NUS-NTU International Joint Research Unit, UMI 3654, Singapore}
\affiliation{National Institute of Education and Institute of Advanced Studies,
Nanyang Technological University, 1 Nanyang Walk, Singapore 637616}

\begin{abstract}

Quantum networks provide a prominent platform for realizing quantum information processing and quantum communication, with entanglement being a key resource in such applications. Here, we describe the dissipative transport protocol for entangled states, where entanglement stored in the first node of quantum network can be transported with high fidelity to the second node via a 1D chiral waveguide. In particular, we exploit the directional asymmetry in chirally-coupled single-mode ring resonators to transport entangled states. For the fully chiral waveguide, Bell states, multipartite $W$-states and and Dicke states can be transported with fidelity as high as $0.954$, despite the fact that the communication channel is noisy. Our proposal can be utilized for long-distance distribution of multipartite entangled states between the quantum nodes of the open quantum network.

\end{abstract}

\maketitle

\section{Introduction}
Quantum networks \cite{kimble2008quantum,reiserer2015cavity} are essential for realizing distributed quantum computing and large scale quantum communication, with entanglement being a key resource in such applications. In this context, the main task and at the same time an outstanding challenge is the high fidelity transfer of quantum states over long distances despite having noise and dissipation present in the quantum channel \cite{nielsen2002quantum}. It is well known that in open quantum systems, dissipation arises as a result of the system coupling with the reservoir \cite{breuer2002theory}, which consequently causes decoherence in the system. In this paper, we show that the dissipative channel can be used for transporting entangled states. In general, long-distance processes in open quantum systems are challenging because of non-Markovian effects due to the non-negligible time delay between the nodes of a quantum network. Non-Markovianity has been shown to be detrimental to both quantum state transfer and entanglement generation between the nodes~\cite{Gonzalez_Ballestero_2013,fang2018non}. 

 %Spin chains review put them down
 %state transfer review put them down
There have been several theoretical proposals \cite{cirac1997quantum,nikolopoulos2014quantum,PhysRevLett.118.133601,dlaska2017robust,stannigel2010optomechanical,stannigel2011optomechanical,yao2013topologically,ramos2016non,zheng2013persistent,van2019long,PhysRevX.7.011035} as well as actual experimental realizations \cite{hofmann2012heralded,ritter2012elementary,rosenfeld2017event} for the quantum state transfer (QST) of a single qubit in quantum optical networks, where fast information transfer is achieved with help of photons  (``flying qubits''). 
%For instance, in the seminal proposal by Cirac \textit{et al.} \cite{cirac1997quantum}, the qubit state is written in a three-level atom and by applying control pulses, the state can then be transferred to the second node using a time-symmetric photonic wave packet, which mimics the reverse process of the wavepacket emission. 
In all the above proposals, there are few demanding requirements that are hard to be met experimentally: external control pulses that have non-trivial temporal shapes (photonic wave packets are required to be time-symmetric), time-dependent cavity-atom and fiber-atom interaction strengths. Moreover, to the best of our knowledge, there are no existing protocols for long-distance entanglement transfer in the optical frequency domain. 

On the other hand, spin chains can alleviate the issue of sensitive control of system parameters and realize quantum systems with minimal control (coupling constants are fixed in time), and entanglement transfer has been demonstrated in several theoretical manuscripts  \cite{bose2007quantum,rafiee2011entanglement,bayat2010entanglement,sousa2014pretty,ji2015quantum,man2014controllable,vieira2018almost,banchi2011long} in Heisenberg-type spin chains. However, these systems can only realize short-distance state transfer, as experimentally one is limited by the number of spins. It is also widely believed that increasing the length of a spin chain will worsen transfer fidelities due to dispersion effects~\cite{banchi2011long}.
%Quantum networks consist of nodes, which are usually formed with atoms. Nodes are then linked together through the quantum channel via photons (referred in this picture as (``flying qubits"). 

%entenglement as a resource
 %It is well known that if Bob shares entangled qubit pairs both with Alice and Carol, then by using classical communication channel, crucial protocols like teleportation \cite{bennett1993teleporting}, entanglement swapping \cite{zukowski1993event,coecke2004logic} and quantum cryptography \cite{bennett2014quantum,ekert1991quantum} can be achieved with high success rate. These breakthrough achievements demonstrate that entanglement is a vital ingredient for realizing quantum information and quantum communication tasks \cite{nielsen2002quantum}. In this Letter we target a rather different question: if Bob, who is located in the first node of quantum network, has a known entangled state, how can he transfer (without using a classical communication channel) it to the Alice, who is located in the second node, with a high fidelity? 
 %by choosing system parameters in advance  to maximize the value of the transfer fidelity. 
 
 %Chiral waveguides chiral route take as motivation connect them with cascaded systems praise markovianity put down non-Markovian effects
Quite remarkably, using chiral waveguides, the merits of quantum optical networks (fast information transfer with ``flying qubits'') and spin chain networks (minimal control over system parameters) can be combined. In quantum optics, chirality arises, for instance,  in atom-waveguide coupled systems when the symmetry of photon emission in the left and right directions is broken \cite{lodahl2017chiral}. This effect appears as a result of spin-orbit coupling, and has been experimentally demonstrated in photonic waveguides \cite{sollner2015deterministic}. Chiral systems have been shown to be fruitful for realizing quantum networks \cite{PhysRevA.91.042116,ramos2016non,mahmoodian2016quantum}. In Ref. \cite{gonzalez2015chiral}, it was argued that the maximum achievable concurrence between two atoms is 1.5 times higher as compared to the non-chiral counterparts.
%with additional benefit that the generated entanglement is insensitive to the distance between the atoms.
 %with the added feature that multipartite entangled states can be generated with almost all-to-all connectivity in the steady state.
 
Interestingly, systems with perfect chirality realize the paradigm of cascaded systems \cite{PhysRevLett.70.2273,gardiner1993driving,carmichael2007statistical}, where two systems are coupled unidirectionally without information backflow.  Cascaded systems, even when separated by long distances, can then be described under the Born-Markov approximation with retardation effects accounted for by a simple redefinition of the time and phase of the second node \cite{carmichael2007statistical}, such that the resulting Markovian master equation does not contain source retardation. Physically, the time delay between the two nodes is not important for cascaded systems since there is no back-action from the target to the source, by definition. Here, we exploit the Markovianity provided by cascaded systems as a suitable platform to achieve high-fidelity entanglement transfer, despite the noise being present in the quantum channel.
%Non-Markovian effects (due to the finite time delay between nodes) were shown to be detrimental for generating entangled states in atom-waveguide systems \cite{Gonzalez_Ballestero_2013,fang2018non}. 
  
  %Ring cavities take zoller as motivation
%To implement functional quantum networks, photonic quantum devices \cite{o2009photonic} are key components and play an important role in realizing light-matter interface. 
%Micro-chip based systems such as microtoroidal and microdisk cavities hold a promise to realize scalable quantum networks \cite{vahala2003optical}. Moreover, by coupling tapered fibre with ring resonator, the high (up to 0.997) coupling efficiency of light in and out of the microtoroidal resonator have been experimentally achieved  \cite{aoki2006observation}. In Ref. \cite{PhysRevLett.118.133601} it was demonstrated that contrary to the direct fiber-atom coupled case, QST mediated by a ring cavity becomes resilient to thermal noise.  

%that by introducing ring cavities in the atom-waveguide system it is possible to implement a quantum state transfer
%protocol in the microwave frequency domain which is immune to the thermal noise if the waveguide is perfectly chiral. 

Motivated by Refs. \cite{gonzalez2015chiral,PhysRevLett.118.133601}, we couple ring cavities with chiral waveguides to obtain unidirectional effective coupling between the cavities. In particular, we benefit  from Markovian dynamics thanks to the cascaded systems naturally arising from the high chirality.  Each node of our quantum network consists of a $N$-particle atomic ensemble which is coupled to the ring cavity. It is important to highlight that ring cavities introduce greater control over the system compared to the bare atom-fiber coupled case, where the transport fidelity is significantly lowered due to all-to-all long-range interactions between the atoms. By suitably optimizing over the system parameters, we demonstrate the transport of maximally entangled Bell states, Dicke states and $W$-states for up to 20 qubits.  For clarity, we remark that the entanglement transport here is not necessarily QST, although the transport of $W$-states and Bell states can be applied to the QST of unknown qubit and qutrit states respectively. 

	Compared to other schemes, our minimal control proposal has various advantages. Firstly, the scheme works in the weak coupling regime with no external driving field required. Also, the optimal transport of entanglement occurs dynamically. This potentially can lead to faster transport compared to steady state schemes \cite{PhysRevA.91.042116}. Moreover, the entanglement transport is not dependent on the distance between the atoms.
	%Our proposal requires minimal control over the system parameters contrary to other proposals which require external pulses with demanding temporal shapes.  
%We highlight, that our proposal requires minimal control over the system parameters contrary to other proposals which require external pulses with demanding temporal shapes. Moreover, since our entanglement is achieved dynamically it is faster compared to steady state counterparts,  and finally our protocol can easily cover large distances thanks to Markovian dynamics in cascaded systems.

%
 
\section{Chiral waveguide-QED system}
\begin{figure}
\subfloat{%
  \includegraphics[width=\linewidth]{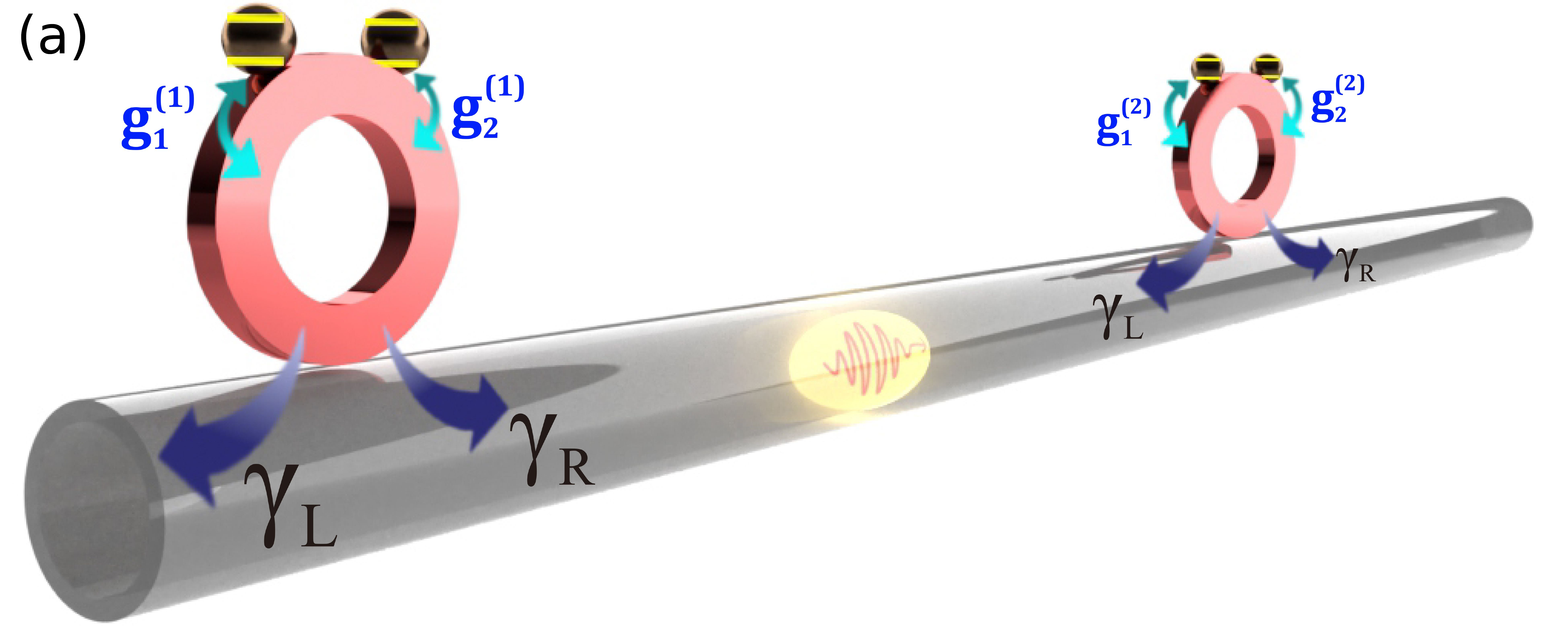}%
  \label{setup}%
}\hfill
\subfloat{%
  \includegraphics[width=.499\linewidth]{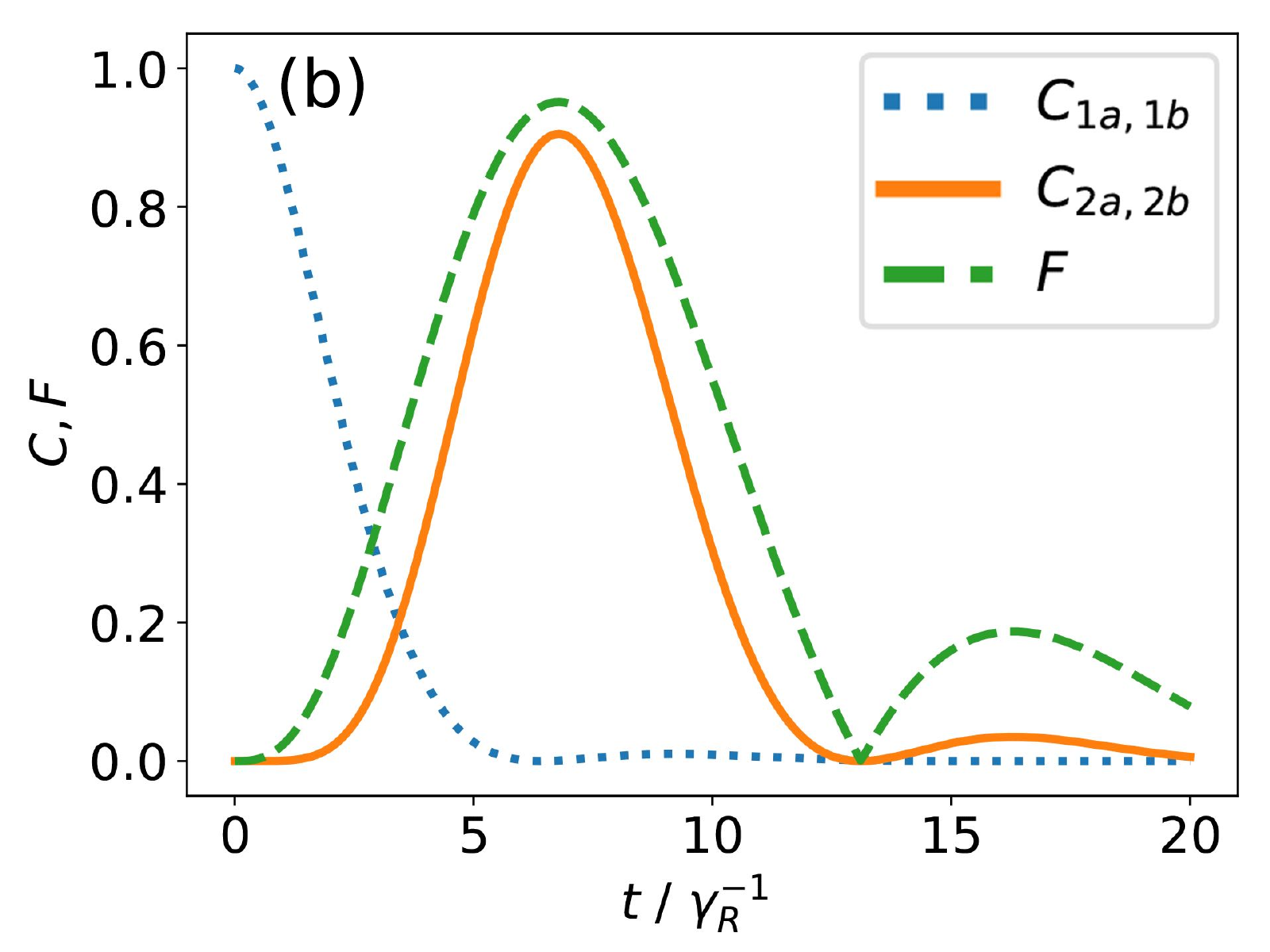}%
  \label{fig:transfer_chiral}%
}\hfill
\subfloat{%
  \includegraphics[width=.499\linewidth]{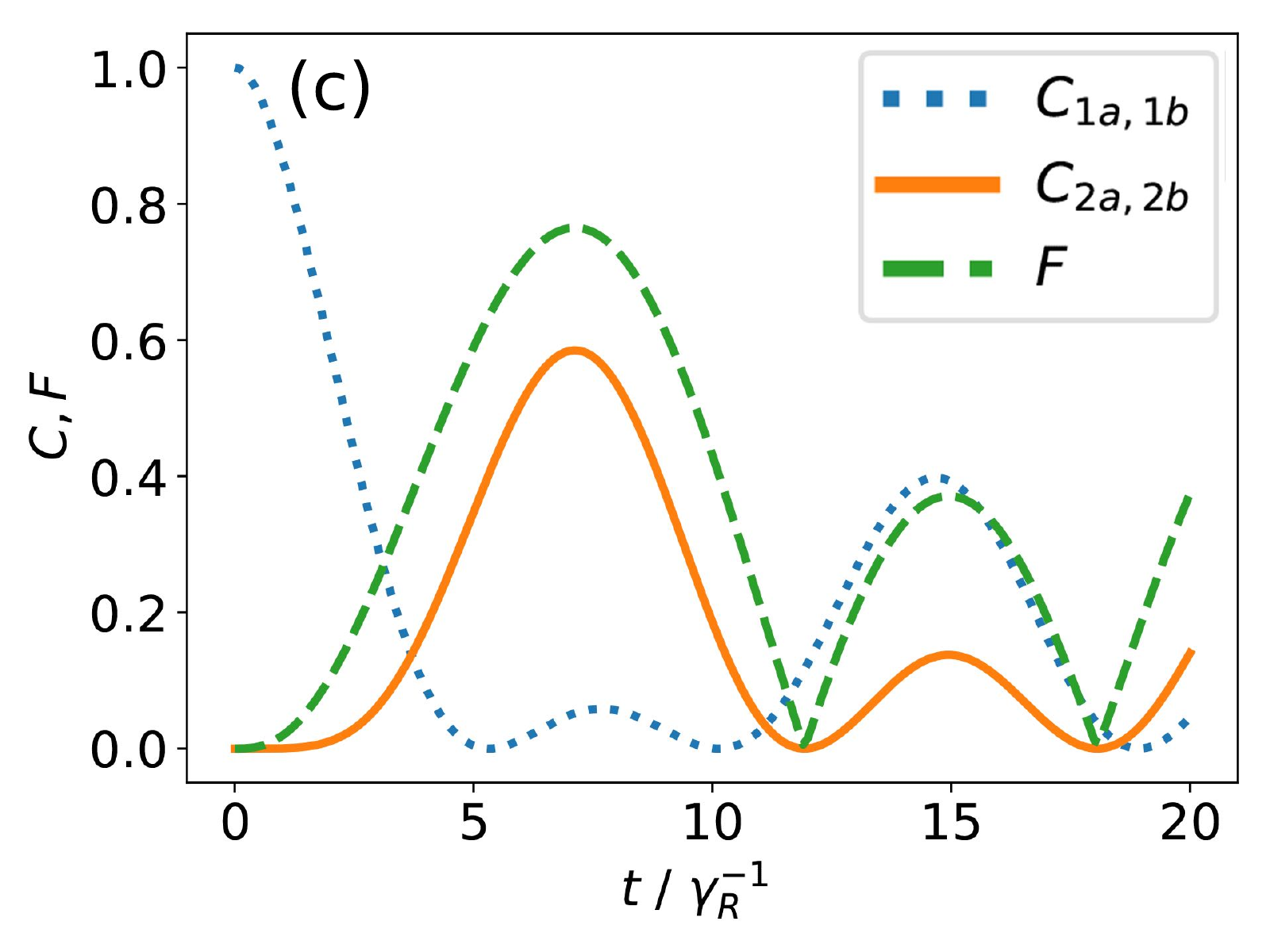}%
  \label{fig:transfer_nonchiral}%
}
	\caption{(a) The proposed setup for entanglement transport. Each node comprises $N$ qubits ($N=2$ in the figure) coupled to a single cavity mode. Chirality is enforced by setting $\gamma_L \neq \gamma_R$. Concurrence and fidelity for the transport of Bell state $\ket{\Psi^+}$, with $C_{\text{1a,1b}}$ and $C_{\text{2a,2b}}$ denoting the qubit concurrence in the left and right nodes respectively. (b) Chiral coupling with $\gamma_L = 0$ (c) Non-chiral coupling with $\gamma_L = \gamma_R$ and $kD = \pi$. Cavity-atom coupling is set at the optimal value $g_1 = g_2 = 0.3\gamma_R$.}
	\label{}
	\end{figure}

The system in consideration comprises two nodes coupled to a 1D waveguide, shown in Fig. \subref*{setup}. Each node comprises $N$ qubits coupled to a single cavity mode where the transition frequencies of the qubits and resonant frequency of the cavity are $\omega_l^{(j)}$ and $\omega_{cj}$ respectively. The atom-cavity coupling strength is given by $g_l^{(j)}$.
The bosonic operators for the cavity mode are $a_j^\dag$ and $a_j$, satisfying the canonical commutation relation $[a_j, a_{j^\prime}^\dag] = \delta_{j j^\prime}$. The waveguide is treated as a common reservoir, with bosonic operators $b_\lambda^\dag (\omega)$ and $b_\lambda (\omega)$ satisfying the commutation relation $[b_\lambda (\omega), b_{\lambda^\prime}^\dag (\omega^\prime)] = \delta_{\lambda \lambda^\prime} \delta(\omega - \omega^\prime)$. The interaction strength between the cavities and the waveguide (at position $x_j$) is characterized by the decay rate $\gamma_{\lambda}$. Here we assume that cavity losses into non-waveguide modes are negligible, which can be realized in photonic waveguides with $\beta$ factors close to unity \cite{PhysRevLett.113.093603} . The spontaneous decay of the qubits is described by an interaction with independent baths at a decay rate $\Gamma_{jl}$, where the first index denotes the cavity and the second index denotes the qubit. The bath operators $c_l^{(j) \dag} (\omega)$ and $c_l^{(j)} (\omega)$ satisfy the commutation relation $[c_l^{(j)} (\omega), c_{l^\prime}^{(j^\prime)\dag} (\omega^\prime)] = \delta_{j j^\prime} \delta_{l l^\prime} \delta(\omega - \omega^\prime)$.

By tracing out the waveguide mode, and applying the Born-Markov approximation, the Lindblad master equation for the system can be found as \cite{PhysRevA.91.042116} (setting $\hbar = 1$, details in Appendix \ref{appendixA})
\begin{equation}
\begin{split}
\dot{\rho} &= -i [H_{\text{eff}}, \rho] + \gamma_L \mathcal{D} [e^{ikx_1} a_1 + e^{ikx_2} a_2] \rho \\
&+ \gamma_R \mathcal{D} [e^{-ikx_1} a_1 + e^{-ikx_2} a_2] \rho + \sum_{j,l} \Gamma_{jl} \mathcal{D}[\sigma_l^{(j)}] \rho
\end{split}
\label{eq:master_eqn}
\end{equation}
with the effective Hamiltonian
\begin{equation}
\begin{split}
H_{\text{eff}} &= \sum_{j,l} [ \omega_l^{(j)} \sigma_l^{(j)\dag} \sigma_l^{(j)}+ \omega_{cj} {a_j}^\dag a_j + g_l^{(j)} ({a_j}^\dag \sigma_l^{(j)} + \text{H.c.}) ] \\
&- i\frac{\gamma_L}{2} (e^{ikD} a_1^\dag a_2 - \text{H.c.}) - i \frac{\gamma_R}{2} (e^{ikD} a_2^\dag a_1 - \text{H.c.}) 
\end{split}
\end{equation}
where $D = |x_2 - x_1|$ is the distance between the nodes. The Lindblad superoperator in the master equation is given by $\mathcal{D}[\hat{O}]\rho =  O \rho O^\dag - \frac{1}{2} \{O^\dag O, \rho \} $. In the following, we will study the transport of entangled qubit states between the nodes mediated by the waveguide. The case of $N=2$ is first presented to illustrate Bell state transport.

\section{Transport of Bell states with chiral couplings}
Here, we exploit the directional asymmetry by using a chiral light-matter interface, with $\gamma_L = 0, \gamma_R \neq 0$ \cite{PhysRevLett.118.133601}. Here, chirality is defined as $\chi \equiv (\gamma_R - \gamma_L)/(\gamma_R + \gamma_L)$, with the perfectly chiral case corresponding to $\chi = 1$. Using chiral couplings, the setup is essentially a cascaded quantum system \cite{PhysRevLett.70.2273} where the first node is coupled to the second node unidirectionally without backflow of information. In this case, the setup we consider can be used to study long-distance entanglement transport despite the Born-Markov approximation used, since retardation effects in a cascaded quantum system is accounted for by a simple redefiniton of the time and phase of the second node \cite{carmichael2007statistical}.

For simplicity, we assume that the qubit decay rates are much smaller than the cavity decay rates and can be neglected, and the nodes are identical, i.e. $\omega_{l}^{(j)} = \omega_0$, $\omega_{cj} = \omega_c$, $g_l^{(j)} = g_j$, for all $j,l \in \{1,2\}$. The qubits in the first node are denoted by $1a,1b$ while the qubits in the second node are denoted by $2a,2b$. We first prepare the qubits $1a,1b$ in the Bell state $\ket{\Psi^+} = \frac{1}{\sqrt{2}} (\ket{eg} + \ket{ge})$, and consider resonant conditions $\omega_c = \omega_0$ with cavity coupling strength $g_1 = g_2 = 0.3 \gamma_R$. For the case of $N=2$, the entanglement of the two-qubit mixed state $\rho$ is measured by the concurrence, which is defined as
	\begin{equation}
C = \max (0, \lambda_1 - \lambda_2 - \lambda_3 - \lambda_4)
	\end{equation}
where $\lambda_i$, $i = 1,2,3,4$ are the eigenvalues of the matrix $\sqrt{ \sqrt{\rho} \tilde{\rho} \sqrt{\rho}}$ in decreasing order, where $\tilde{\rho} = (\sigma_y \otimes \sigma_y) \rho^* (\sigma_y \otimes \sigma_y)$ is the spin-flipped state \cite{PhysRevLett.80.2245}.

	\begin{figure}
\subfloat{%
  \includegraphics[width=0.499\linewidth]{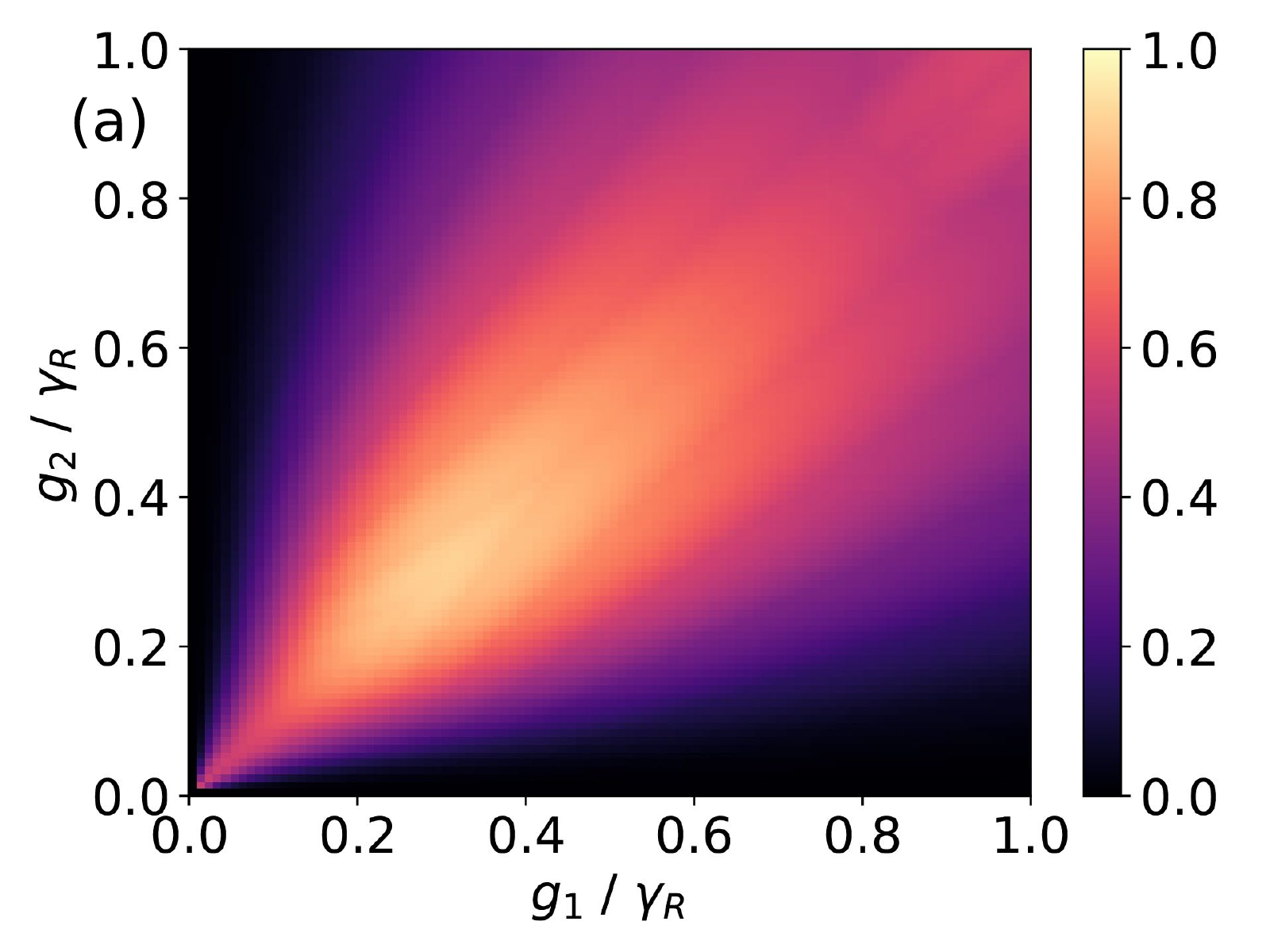}%
  \label{fig:max_g1g2}%
}\hfill
\subfloat{%
  \includegraphics[width=.499\linewidth]{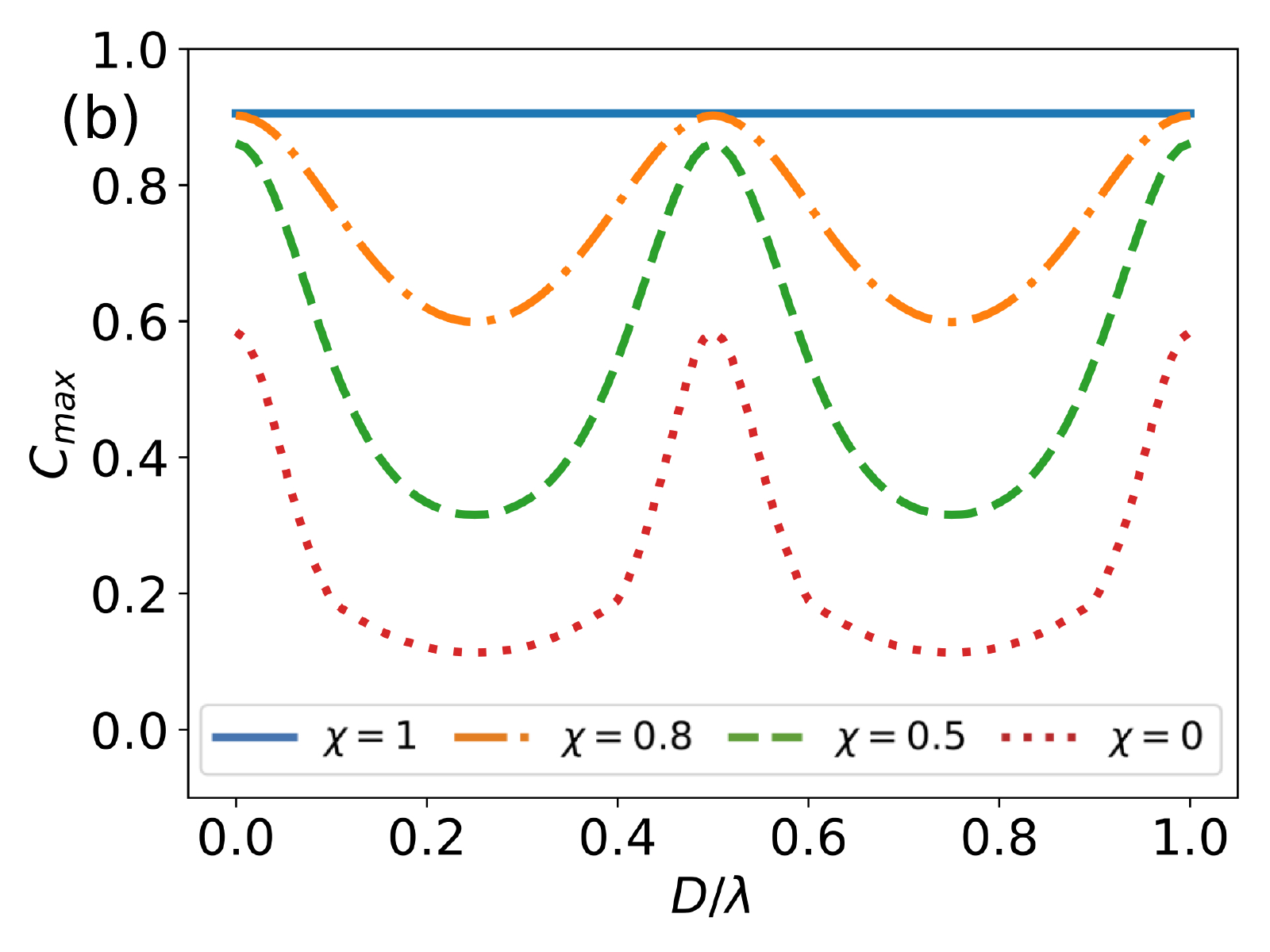}%
  \label{fig:max_chirality}%
}\hfill
\subfloat{%
  \includegraphics[width=.493\linewidth]{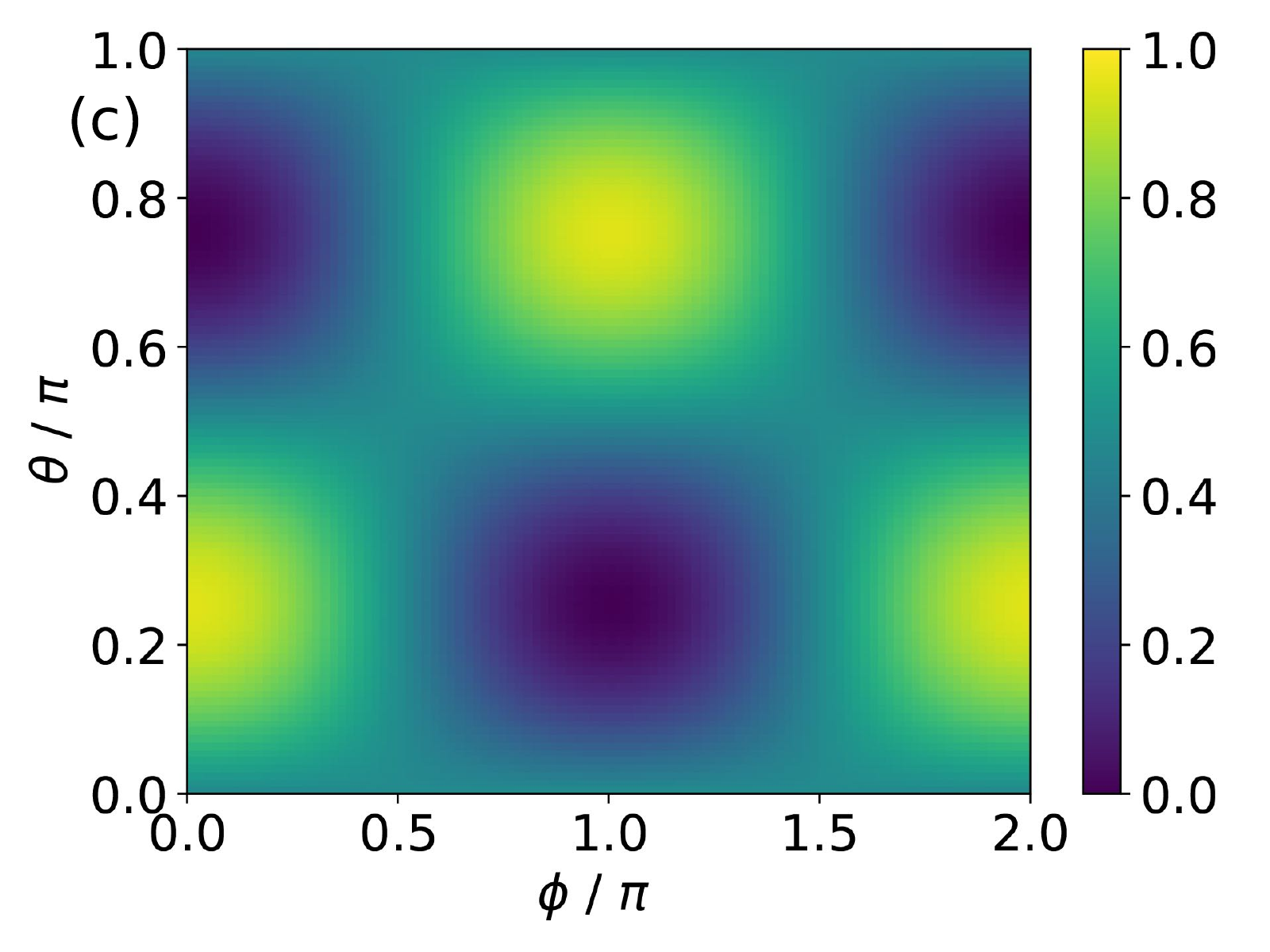}%
  \label{fig:max_angles}%
}\hfill
\subfloat{%
  \includegraphics[width=.493\linewidth]{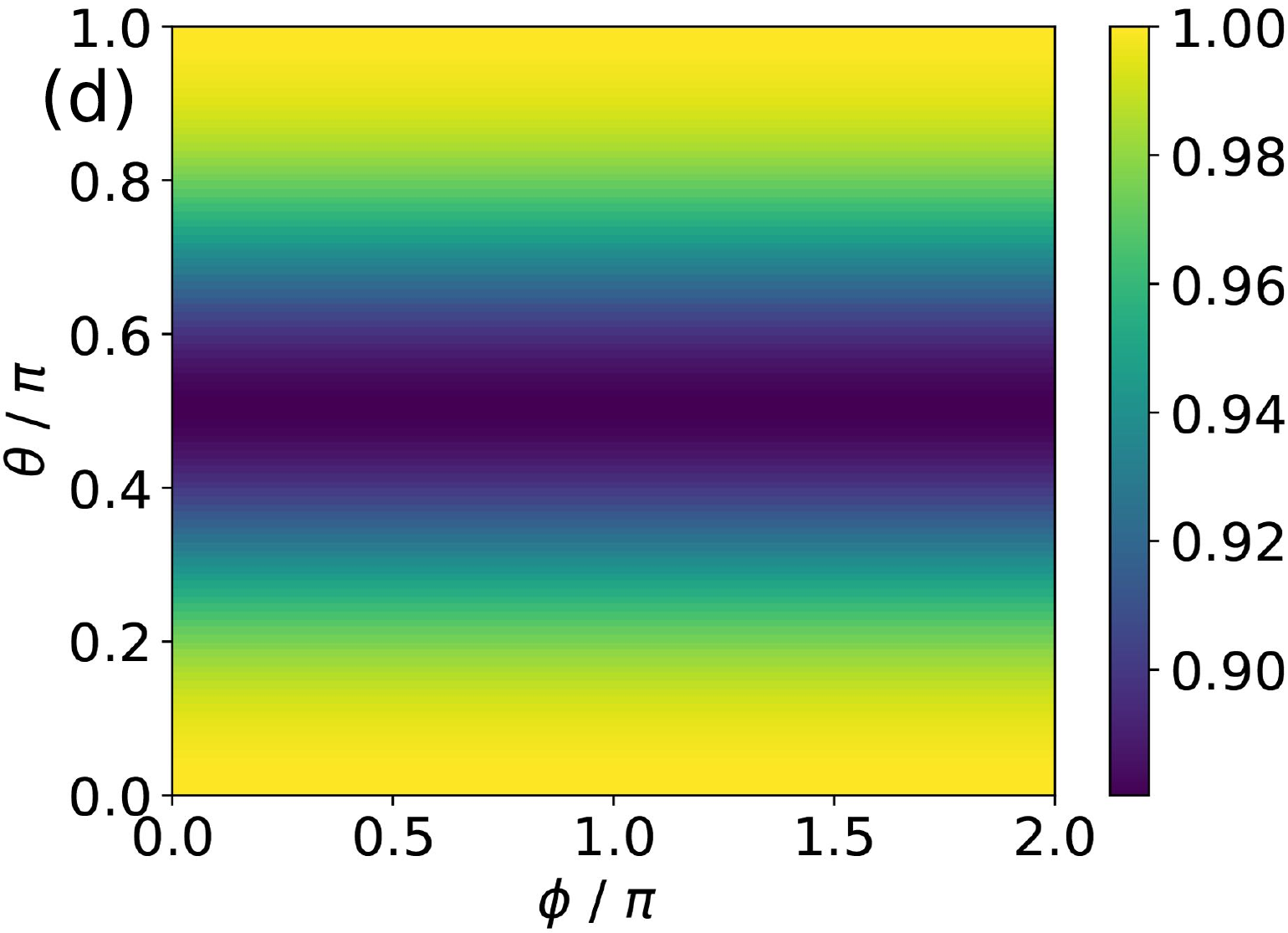}%
  \label{fig:max_angles2}
}\hfill
	\caption{(a) Maximum concurrence of $2a,2b$ against $g_1,g_2$ shows optimal point $g_1 = g_2 = 0.3\gamma_R$. (b) Maximum concurrence of $2a,2b$ against inter-nodal distance $D$. (c) Maximum fidelity of $2a,2b$ against various initial states $\ket{\psi_\alpha}$. (d) Maximum fidelity of $2a,2b$ against various initial states $\ket{\psi_\beta}$.  Other parameters are: $g_1 = g_2 = 0.3\gamma_R$, $\Gamma_{jl} = 0$.}
	\label{}
	\end{figure}

As shown in Fig. \subref*{fig:transfer_chiral}, the concurrence of $1a,1b$ decreases to nearly zero at some time, while concurrence of $2a, 2b$ rises from zero to a maximum of around 0.91. The state fidelity of $2a,2b$ compared to the initial Bell state is around 0.954. This shows that a good entanglement transport can be accomplished. For the case of perfect chirality ($\chi = 1$), due to the cascaded nature of the system, this result is independent of the distance $D$ between the qubits. For the non-chiral case in Fig. \subref*{fig:transfer_nonchiral} where $\gamma_L = \gamma_R$, the maximum concurrence is only around 0.58, even with the optimal distance of $kD = \pi$, where $k$ is the wavenumber of the photon $k = 2\pi / \lambda$ with $\lambda$ the corresponding wavelength. Comparing the fidelity of the qubit state of $2a,2b$ (denoted $\rho_{2}(t)$) with the initial entangled state of $1a,1b$ (denoted $\rho_{1}(0)$) such that $\mathcal{F} = \left(\Tr \sqrt{\sqrt{\rho_{1}(0)} \rho_{2}(t) \sqrt{\rho_{1}(0)}} \right)^2$, Fig. \subref*{fig:transfer_chiral} shows that the maximum fidelity transported, $\mathcal{F}_{\text{max}}$, is around 0.951 (green dashed line), a significant improvement over the non-chiral case in Fig. \subref*{fig:transfer_nonchiral} which gives $\mathcal{F}_{\text{max}} \approx 0.78$. Thus, chiral coupling drastically improves the entanglement transport between the nodes. 

To find the optimal coupling $g_1 = g_2 = g$, we plot the maximum transported concurrence $C_{\text{max}}$ of $2a,2b$ against $g_1$ and $g_2$. As shown in Fig. \subref*{fig:max_g1g2}, the transported concurrence is maximal $(C_{\text{max}} \approx 0.905)$ around $g_1 = g_2 \approx 0.3\gamma_R$. Intuitively, for small couplings, the entanglement does not transport effectively to the cavity, thus the transport is weak. For strong couplings however, the Rabi oscillations between the cavity and the qubits become more significant, which is detrimental to the transport of entanglement via the waveguide. It can also be seen from Fig. \subref*{fig:max_g1g2} that $g_1 = g_2$ is an optimal condition for good entanglement transport.

To illustrate the effect of chirality on the transport, we compare the maximum transported concurrence for different chirality. We comment that although non-Markovian effects should in general be taken into account if one considers long distances with imperfect chirality, this is not required as long as the entanglement is transported much faster than the timescale for information backflow to occur. For example, if the distance between the nodes are such that the time delay $\tau \gg \gamma_R^{-1}$, then the non-Markovian effects do not appear at the much shorter system timescale. As a result, the decay rate $\gamma_L \neq 0$ can be simply regarded as additional leakage of excitation from the second node. Our results indicate that such conditions can be easily achieved for sufficiently long waveguides. Fig. \subref*{fig:max_chirality} shows the comparison for different chirality. For the fully chiral waveguide ($\chi = 1$), $C_{\text{max}}$ is independent of the inter-nodal distance $D$, as previously mentioned. This is simply due to the cascaded nature of the setup. However, when $\gamma_L \neq 0$, $C_\text{max}$ depends on the distance between the nodes. The peak at $D = 0.5 \lambda$ is a result of the spatial localization of the photon wavefunction between the nodes \cite{Gonzalez_Ballestero_2013}, resulting in less detrimental scattering effects which contributes to excitation leakage. The sensitivity of $C_{\text{max}}$ to fluctuations around this `sweet spot' decreases as $\chi$ gets closer to 1. In general, the entanglement transport worsens with decreasing chirality. Intuitively, this can be due to two factors: (i) leakage of excitation from the first node through the left port via $\gamma_L$, which decreases the probability of the second node being excited; (ii) information backflow from the second node back to the first node, which can be detrimental to the transport process. Thus, using chirality, both problems can be addressed simultaneously, leading to good entanglement transport.

Next, we look at the maximum transported fidelity with different initial states of $1a,1b$. To this end, we prepare the qubits in system 1 in the state
	\begin{equation}
\ket{\psi_\alpha} = \cos \theta \ket{eg} + e^{i\phi} \sin \theta \ket{ge}, \quad \theta \in [0, \pi], \phi \in [0,2 \pi]
	\end{equation}
while the qubits in system 2 are initially in the ground state. The cavities are all in the vacuum state initially. From Fig. \subref*{fig:max_angles}, the maximum transported fidelity $(\mathcal{F}_{\text{max}} = 0.951)$ occurs near $\phi =0, \theta = \pi/4$ which corresponds to the Bell state $\ket{\Psi^+}$. The case of $\mathcal{F}_{max} = 0$ occurs near $\phi = \pi, \theta = \pi/4$ which corresponds to the Bell state $\ket{\Psi^-}$. This is because $\ket{\Psi^-}$ is a dark state, and thus does not decay with time. However, $\ket{\Psi^-}$ can be easily transported with the same fidelity of $\mathcal{F}_{\text{max}} = 0.951$ by imposing an $\pi$ phase difference between the two cavity-qubit couplings in the same node, i.e. $g_{1}^{(j)} = - g_{2}^{(j)}$. We also consider the initial state
\begin{equation}
\ket{\psi_\beta} = \cos \theta \ket{gg} + e^{i\phi} \sin \theta \ket{ee}, \quad \theta \in [0, 2\pi], \phi \in [-\pi,\pi]
\end{equation}
with the Bell states $\ket{\Phi^+} = \frac{1}{\sqrt{2}} (\ket{gg} + \ket{ee})$ and $\ket{\Phi^-} = \frac{1}{\sqrt{2}} (\ket{gg} - \ket{ee})$. As shown in Fig. \subref*{fig:max_angles2}, the maximum transported fidelity is independent of $\phi$. The transported fidelity $\mathcal{F}_{\text{max}} \approx 0.954$ at $\theta = \pi/4, \phi=0$ and $\theta = \pi/4, \phi=\pi$ corresponds to the Bell states $\ket{\Phi^{\pm}}$ respectively. The lowest $\mathcal{F}_{\text{max}} \approx 0.88$ occurs at $\theta = \pi / 2$ which is reasonable since that corresponds to the case of transporting a two-excitation state $\ket{ee}$, and the probability of excitation leakage via dissipation to the right waveguide port increases when transferring higher-excitation states. Overall, we have shown that good transport of entanglement is possible for all the Bell states.

\section{Transport of multipartite entanglement with chiral couplings}
\begin{figure}
\subfloat{%
  \includegraphics[width=0.499\linewidth]{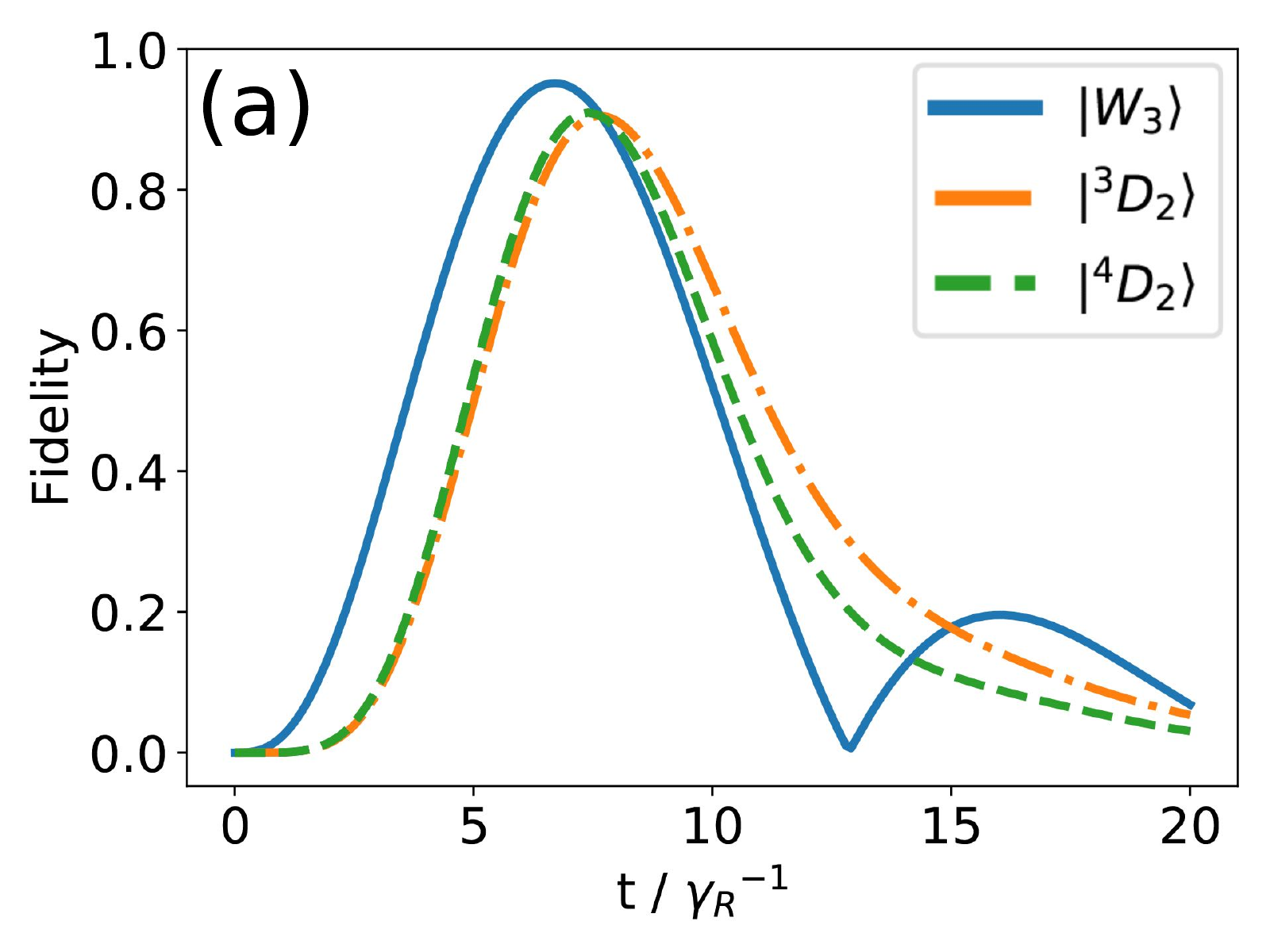}%
  \label{fig:dickestate_transfer}%
}\hfill
\subfloat{%
  \includegraphics[width=0.499\linewidth]{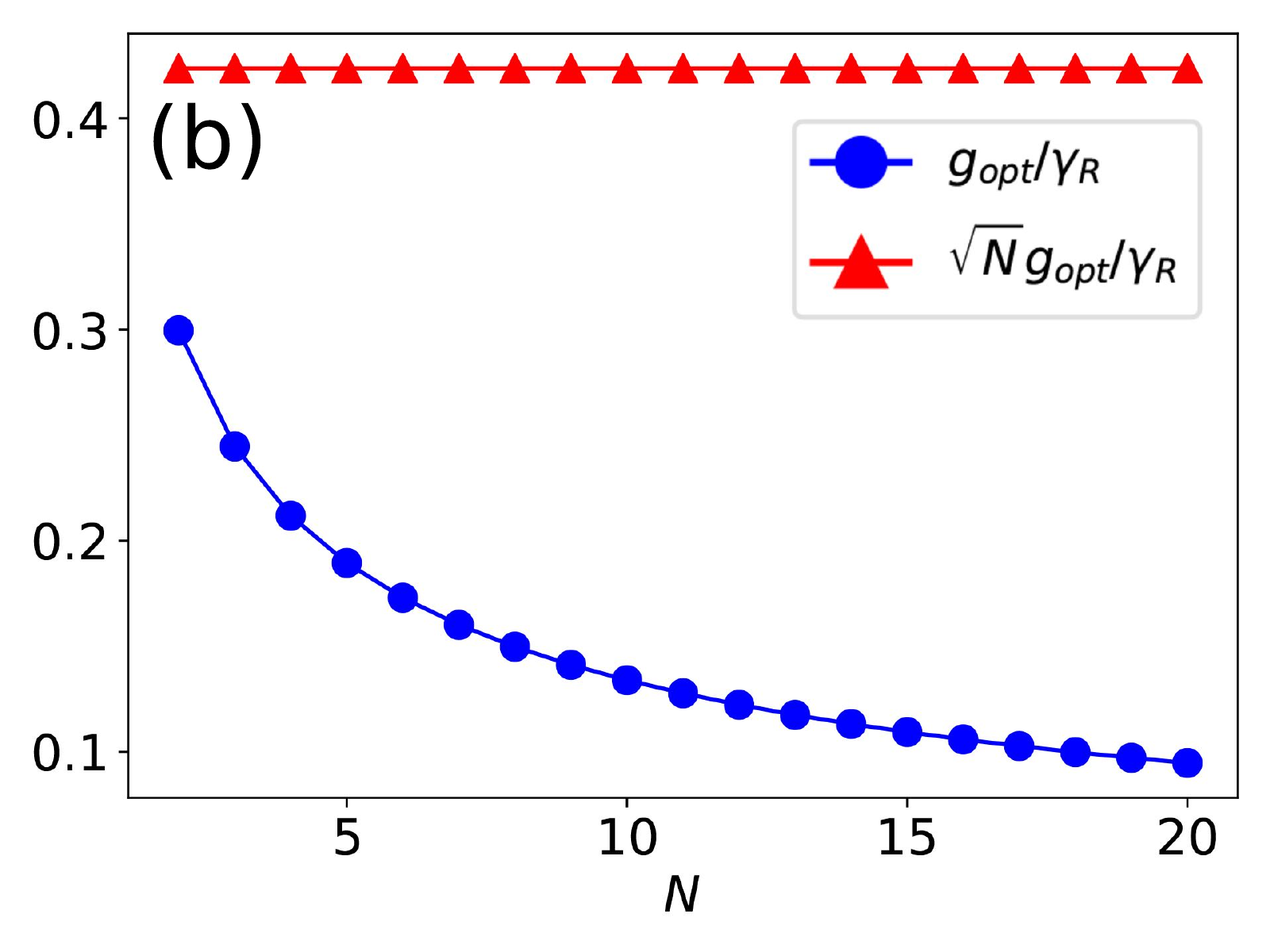}%
  \label{fig:optimal_g}%
}\hfill
\subfloat{%
  \includegraphics[width=0.499\linewidth]{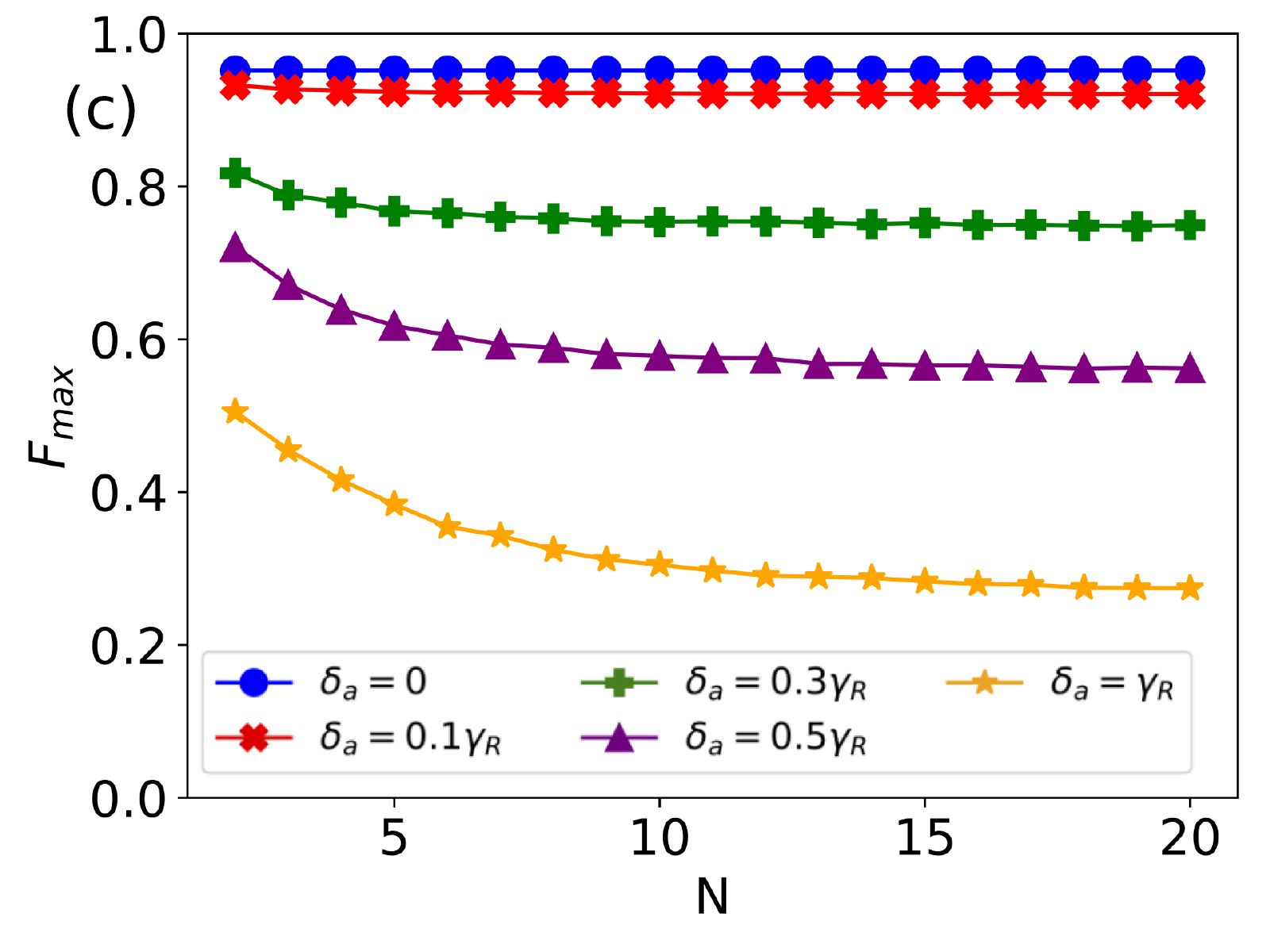}%
  \label{fig:atomdetuning}%
}\hfill
\subfloat{%
  \includegraphics[width=0.499\linewidth]{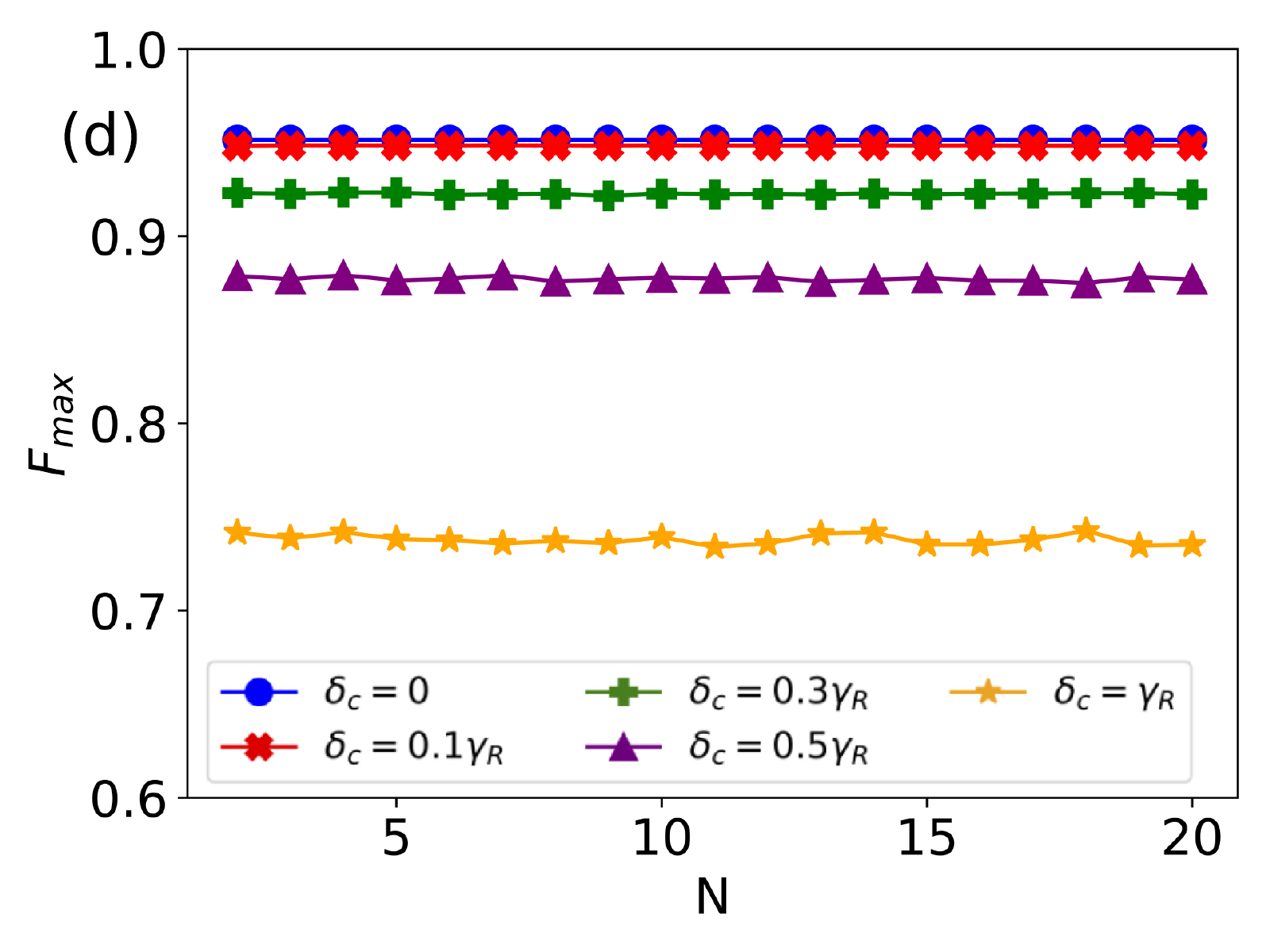}%
  \label{fig:cavitydetuning}%
}\hfill
	\caption{(a) Transport of Dicke states $\ket{W_3}$ $(g_1 = g_2 = 0.248\gamma_R)$, $\ket{^{3} D_2}$ $(g_1 = g_2 = 0.248\gamma_R)$ and $\ket{^{4}D_2}$ $(g_1 = g_2 = 0.215\gamma_R)$. (b) Optimal $g_1=g_2$ for $\ket{W_N}$. (c) Effect of random fluctuations in qubit frequencies on fidelity. (d) Effect of random fluctuations in cavity frequencies on fidelity. Other parameters are: $\gamma_L = 0$, $\Gamma_{jl} = 0$.}
	\label{fig:threequbit_transfer}
	\end{figure}

Here, we demonstrate a generalisation of the entanglement transport scheme, by using $N$ qubits per node. Intuitively, it is clear that states with permutation invariance and low excitations can be transported well using this scheme. In Fig. \subref*{fig:dickestate_transfer} we show the transport of Dicke states denoted by $\ket{^{N}D_{k}}$ which is an equal superposition of $k$ excitations over $N$ qubits. It can be seen that the three-qubit $W$-state $\ket{W_3} = \ket{^{3} D_1}$ is transported with a fidelity of $\mathcal{F}_{\text{max}} \approx 0.954$, while the two-excitation states $\ket{^{3}D_2}$ and $\ket{^{4} D_2}$ are transported with a lower fidelity of 0.905 due to increased leakage of excitation from the second node of the quantum network. $W$-states are extremely useful for quantum information  and communication applications as they are more robust states for encoding single qubit states. Moreover, $W$-states have the unique property (contrary to say, GHZ states) that even if one particle is lost, the rest of $N-1$ qubits will remain in the entangled state. In Fig. \subref*{fig:optimal_g}, the optimal $g_j$ for the transport of $\ket{W_N}$ is plotted. We numerically show that the optimal transport condition for any $\ket{W_N}$ is given by $\sqrt{N} g_{\text{opt}} / \gamma_R \approx 0.43$. In fact, any $\ket{W_M}$ can be mapped onto any $\ket{W_N}$ ($M \neq N$ in general) with the same fidelity of 0.954 as long as this condition is satisfied on each node. To study the effects of detuning on $W$-state transport, we add a random fluctuation $\Delta_a \in [-\delta_a, \delta_a] $ to the qubit frequencies in Fig. \subref*{fig:atomdetuning} and $\Delta_c \in [-\delta_c , \delta_c] $ to the cavity frequencies in Fig. \subref*{fig:cavitydetuning}. The result shows that while the scheme is more robust against fluctuations in cavity frequencies, good transport can also be achieved for $\delta_a \leq 0.1 \gamma_R$.

\section{Role of imperfections} 
The analysis in the previous sections neglected qubit losses by assuming that the decay rate of the cavity is much larger than that of the qubit decay rates. Here, we look at the entanglement transport with qubit losses. Specifically, we prepare the initial state of $1a,2a$ in the Bell state $\ket{\Psi^+}$ and set all qubit decay rates to be equal ($\Gamma_{jl} = \Gamma$) for simplicity. The optimal case (from optimization over system parameters) is presented for each value of $\Gamma$. In general, for larger $\Gamma$, the probability of spontaneous emission of the initially excited qubits in the first node increases. Thus, in order to transport the excitation effectively, the excitation should be transferred to the ring cavity before significant qubit decay occurs, resulting in larger optimal coupling $g_{\text{opt}}$.

\begin{figure}
\subfloat{%
  \includegraphics[width=0.499\linewidth]{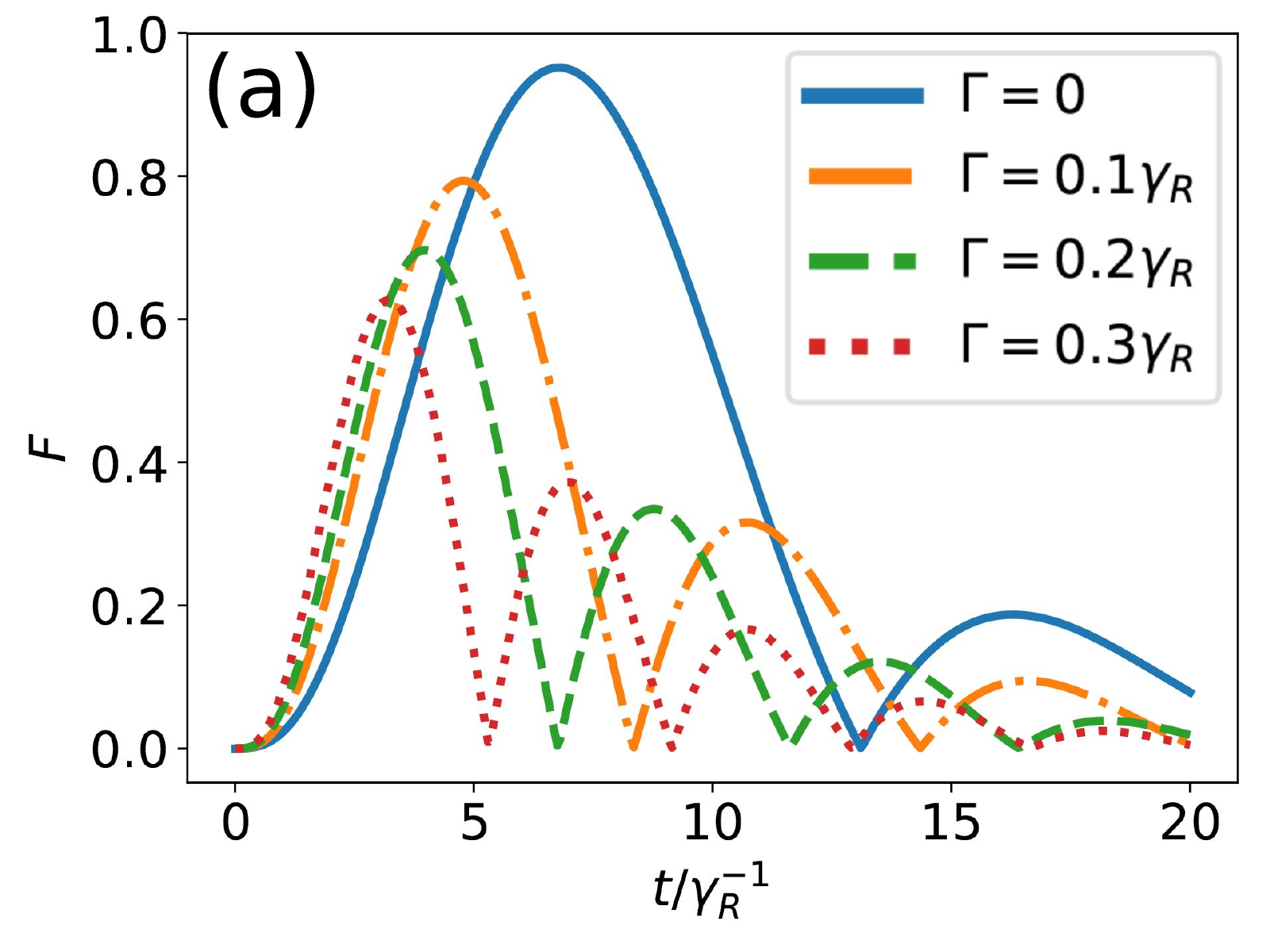}%
  \label{fig:transfer_lossy}%
}\hfill
\subfloat{%
  \includegraphics[width=.499\linewidth]{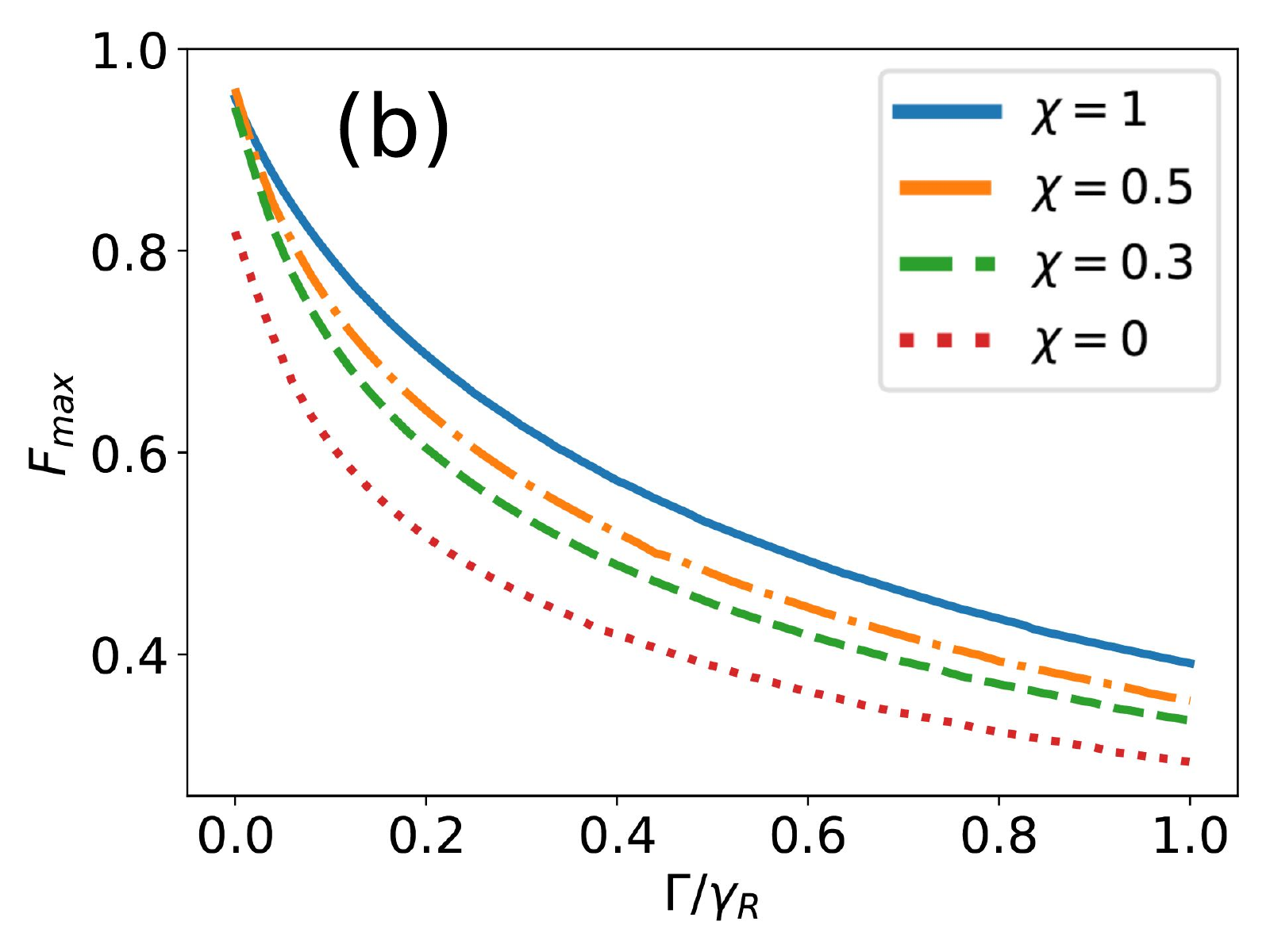}%
  \label{fig:transfer_lossy}%
}\hfill
	\caption{Effects of qubit losses on entanglement transport of $\ket{\Psi^+}$. (a) Fidelity of $2a,2b$ with chiral coupling. (b) Maximum fidelity of $2a,2b$ against qubit decay rate. The optimised case is shown for each value of $\Gamma$.}
\label{}
\end{figure}
Increasing the qubit decay rate, the fidelity decreases as shown in Fig. \subref*{fig:transfer_lossy}. Here, we set the inter-nodal distance to be at the `sweet spot' $kD = \pi$. A comparison between the chiral ($\chi = 1$) and non-chiral ($\chi = 0$) cases shows that as long as the qubit decay rate is within $\Gamma < 0.1 \gamma_R$, the chiral system remains advantageous over the ideal non-chiral case in terms of entanglement transport. As mentioned earlier, the entanglement transport at $kD = \pi$ is also relatively insensitive to small imperfections in chirality. Thus, perfect chirality is not required for the transport scheme to work well. 

\section{Proposal for experimental implementations}
Our scheme can be experimentally realized in two different platforms. Firstly, the authors in Ref. \cite{PhysRevA.100.053841} have recently established a photonic interface of chiral cavity QED using the coupled photonic crystal (PC) and plasmon nanoparticle structure. In this system, the rate of circular polarized photons emitting into the PC waveguide is one order of magnitude larger than that without the nanoparticle, with about 95 \% of photons propagating unidirectionally along the waveguide. In particular, a silver nanoparticle which serves as a nanocavity, is embedded inside the W1 PC where the electric field is mainly located with antisymmetric local helicity distribution. The efficient chiral coupling is controlled by the relative position between the emitter and silver nanoparticle. By applying a strong magnetic field and suitably choosing the laser frequency, two circular polarized states of the atoms or quantum dots can be generated to emit only clockwise or anti-clockwise photons \cite{sollner2015deterministic}.

Another possibility to realize our proposal is by coupling a single $^{85}$Rb atom with a micro-bottle resonator, which is interfaced with a silica nanophotonic tapered fiber. As experimentally demonstrated in Refs. \cite{PhysRevX.5.041036, PhysRevLett.110.213604}, chirality naturally arises in such a system due to the strong transverse confinement of the light in the micro-bottle resonator which results in different circular polarizations for the clockwise and anti-clockwise TM modes in the WGM microresonator. The selection rules for dipole transitions prevent the atom from emitting light into the counter propagating TM mode. Consequently, the chirality arises because the atom has different interaction cross-sections (up to one order of magnitude difference) for the two orthogonal circular polarizations. For our proposal, it suffices to initialize the atoms in the second node in the outermost $m_F = 3$ Zeeman sublevel of the $F=3$ hyperfine ground state. The excited state corresponds to the $F^{'} = 4, m_F = 4$ state. Similarly, we can prepare the ensemble of atoms in the first node in the entangled state. A magnetic field of $B = 4.5$G can be applied along the resonator axis to lift the degeneracy.

\section{Conclusion} 
In this paper, we have proposed a protocol for transporting entanglement between the two nodes in open quantum network, where we demonstrated that dissipation can be useful to achieve the task, contrary to the common notion that dissipation creates decoherence. By coupling ring cavities with a chiral 1D waveguide, we demonstrate entanglement transport, with the entangled state stored in the atomic ensembles which are coupled to the ring cavities. Consequently, our scheme can be implemented experimentally, by coupling a photonic waveguide with ring cavities which are then coupled to an atomic ensemble, to realize the elementary unit of a quantum node. The communication channel is realized by the chiral waveguide. We have found optimal system parameters for the transport of maximally entangled Bell states and for up to 20-qubit $W$-states. As an application of our results, the quantum transport of $W$-states and Bell states can be exploited to achieve QST of unknown qubit and qutrit states respectively. We highlight that our proposal requires minimal control over the system parameters contrary to other proposals which require external pulses with demanding temporal shapes and time-dependent cavity couplings \cite{cirac1997quantum,nikolopoulos2014quantum,PhysRevLett.118.133601,dlaska2017robust,stannigel2010optomechanical,stannigel2011optomechanical,yao2013topologically,ramos2016non,zheng2013persistent,van2019long}. Moreover, since the entanglement transport is achieved dynamically it is faster compared to its steady state counterparts, which requires timescales on the order of $10^2 \gamma_{R}^{-1}$ \cite{PhysRevA.91.042116,matsuzaki2018,PhysRevA.99.032348}. Finally, our protocol can easily be applied to long-distance transport by utilising the Markovianity in cascaded systems. This can potentially be significant for the efficient distribution of entanglement within a quantum network.

\section*{Acknowledgments}
D. A. and W.K. M. would like to acknowledge Marc-Antoine Lemonde for helpful discussions and feedback. The authors thank Jingu Pang for the waveguide diagram. The IHPC A*STAR Team would like to acknowledge the support from the National Research Foundation Singapore (Grants No. NRF2017NRFNSFC002-015, No. NRF2016-NRF-ANR002, No. NRF-CRP 14-2014-04) and A*STAR SERC (Grant No. A1685b0005). D. A., L.C. K. and J. Y. acknowledges support from National Research Foundation Singapore (Grant No. 2014NRF-CRP002-042).

\bibliographystyle{apsrev4-1}
\bibliography{Bibliography}

\onecolumngrid
\appendix
\section{Derivation of the effective master equation}
\label{appendixA}

In this Appendix, we derive the effective master equation from tracing out the degrees of freedom of the common bath, which in this case is the 1D waveguide. Note that this derivation is similar to the approach taken in \cite{PhysRevLett.118.133601}. For simplicity, we neglect qubit decays here and assume that the ring resonators are at a common frequency $\omega_c$. The Hamiltonian is given by $H = H_B + H_S + H_{SB}$, where
\begin{equation}
\begin{split}
H_S &= \sum_{j=1}^2 \sum_{l=1}^N [\omega_l^{(j)} \sigma_l^{(j) ^\dag} \sigma_l^{(j)} + \omega_{cj} a_j^\dag a_j + g_l^{(j)} (a_j^\dag \sigma_l^{(j)} + \text{H.c.}) ]\\
H_{B} &= \sum_{\lambda = L,R} \int d\omega \, \omega b_\lambda^\dag (\omega) b_\lambda (\omega) +\sum_{j=1}^2 \sum_{l=1}^N \int d\omega \, \omega c_l^{(j) \dag} (\omega) c_l^{(j)} (\omega) \\
H_{SB} &= i \sum_{j=1}^2 \sum_{\lambda=L,R} \int d\omega \, \sqrt{\frac{\gamma_{\lambda}}{2\pi}} \bigg( b_\lambda^\dag (\omega) e^{-ikx_j} a_j - \text{H.c.} \bigg) + i \sum_{j=1}^2 \sum_{l=1}^N \int d\omega \, \sqrt{\frac{\Gamma_{jl}}{2\pi}} \bigg(c_{l}^{(j) \dag} (\omega) \sigma_l^{(j)} - \text{H.c} \bigg)
\end{split}
\label{eq:hamiltonian}
\end{equation} 
Setting $\Gamma_{jl} = 0$ from the Hamiltonian in Eq. (\ref{eq:hamiltonian}), and choosing a frame rotating with the cavity and bath, i.e. $U = \exp [i( {\sum_j \omega_{cj} a_j^\dag a_j + \sum_\lambda \int d\omega \,\omega b_\lambda^\dag (\omega) b_\lambda (\omega)} )]$ and applying the transformation $H = UHU^\dag - i\dot{U}U^\dag$, we have
\begin{equation}
\tilde{H}_{SB}(t) = i \sum_{\lambda, j} \int d\omega \sqrt{\frac{\gamma_\lambda}{2\pi}} \bigg( b_\lambda^\dag (\omega) a_j e^{i (\omega - \omega_{c})t} e^{-i\omega x_j / v} - e^{i (\omega - \omega_{c})t }e^{i\omega x_j / v} a_j^\dag b_\lambda (\omega) \bigg)
\end{equation}
From Heisenberg equations of motion, we have
\begin{equation}
\dot{b}_\lambda (\omega,t) = i [ H, b_\lambda (\omega,t)] = \sum_{j=1,2} \sqrt{\frac{\gamma_\lambda}{2\pi}} a_j (t) e^{i (\omega - \omega_{c}) t} e^{-i \omega x_j / v}
\end{equation}
which can be formally integrated to obtain
\begin{equation}
b_\lambda (\omega, t) = b_\lambda( \omega, 0) + \int_0^t ds \sum_j \sqrt{\frac{\gamma_\lambda}{2\pi}} a_j (s) e^{i (\omega - \omega_{c})s} e^{-i\omega x_j / v}
\end{equation}
For an arbitrary system operator $X(t)$, the Heisenberg equation reads
\begin{equation}
\dot{X}(t) = \sum_{\lambda, j} \int d\omega \sqrt{\frac{\gamma_\lambda}{2\pi}} ( b_\lambda^\dag (\omega,t) e^{i (\omega - \omega_{c})t} e^{-i \omega x_j / v} [X(t), a_j (t)] - b_\lambda (\omega,t) e^{-i (\omega - \omega_{c})t} e^{i \omega x_j / v} [X(t), a_j^\dag (t)] )
\end{equation}
Substituting $b_\lambda (\omega,t)$ into $\dot{X} (t)$ and defining $b_\lambda (t) \equiv \frac{1}{\sqrt{2\pi}} \int d\omega b_\lambda (\omega) e^{-i(\omega - \omega_j)t }$ and $k = \omega_0 / v$, we have
\begin{equation} 
\begin{split}
\dot{X}(t) &= \sum_{\lambda, j} \sum_{\lambda, j} \sqrt{\gamma_\lambda} b_\lambda^\dag (t - x_j / v) e^{-i k x_j} [X(t), a_j (t)] - [X(t), a_j^\dag (t)] b_\lambda (t - x_j /v) e^{ikx_j} \\
&+ \sum_{\lambda, j, l} \frac{\gamma_\lambda}{2\pi} \int_0^t ds \int d\omega e^{i (\omega - \omega_c) (t-s)} e^{-i \omega x_{jl} / v} a_l^\dag (s) [X(t), a_j (t)] - e^{-i (\omega - \omega_c) (t-s)} e^{i \omega x_{jl} / v} a_l (s) [X(t), a_j^\dag (t)]
\end{split}
\end{equation}
We can perform the Born-Markov approximation by treating the time delay $x_{jl} / v$ between the two atoms to be very small. Thus,
\begin{equation} 
\begin{split}
\sum_{l} \frac{1}{2\pi} \int_0^t ds \int d\omega e^{i (\omega - \omega_j) (t-s)} e^{-i \omega x_{jl} / v} a_l^\dag (s) &= \sum_l \int_0^t ds \delta(t - x_{jl}/v -s) e^{-ikx_{jl}} a_l^\dag (s) \\
&\approx \frac{1}{2} a_l^\dag (t) + \sum_l \theta(x_{jl}/v) e^{-ikx_{jl}} a_l^\dag (t)
\end{split}
\end{equation}
where the first term is the contribution from $x_{jl}/v < 0$ and the second term is from $x_{jl}/v > 0$. The Markov approximation is also applied to the second term $\sigma_l^\dag (t - x_{jl}/v) \to \sigma_l^\dag (t)$.
Next, we substitute this into the equation for $\dot{X}(t)$ and take averages. Since the bath is initially in the vacuum, $\braket{b_\lambda (t)} = 0$. Thus,
\begin{equation}
\begin{split}
\braket{\dot{X}(t)} &= \sum_{\lambda j} \frac{\gamma_\lambda}{2} \bigg( \braket{ a_j^\dag(t) [X(t), a_j (t)]} - \braket{ [X(t),a_j^\dag (t)] a_j (t)} \bigg) \\
&+ \sum_{\lambda j l, x_j > x_l} \gamma_\lambda \bigg( e^{-ikx_{jl}} \braket{a_l^\dag (t) [X(t), a_j (t)]} - e^{ikx_{jl}} \braket{ [X(t), a_j^\dag (t)] a_l(t)} \bigg)
\end{split}
\end{equation}
To obtain the master equation, we first note that the average is the same in both Schr\"{o}dinger picture and Heisenberg picture, thus $\Tr ( X(t) \rho(0) ) = \Tr (X \rho(t))$, that is, we can move the time dependence from system operator to density operator. For example, the first term on the RHS can be written as 
\begin{equation}
\braket{a_j^\dag (t) [X(t), a_j (t)] } = \Tr ( a_j^\dag X a_j \rho(t) - a_j X a_j^\dag \rho(t) ) = \Tr ( X [a_j, \rho(t) a_j^\dag])
\end{equation}
using the cyclic property of trace. Doing this for all the terms and noting that the equation holds for all $X(t)$, we have
\begin{equation}
\dot{\rho} (t) = -i [H_S, \rho(t)] + \sum_{\lambda j} \frac{\gamma_\lambda}{2} ( [a_j, \rho(t) a_j^\dag] - [a_j^\dag, a_j \rho(t)] ) + \sum_{\lambda j l, x_j > x_l} \gamma_\lambda \bigg( e^{-ikx_{jl}} [\sigma_j, \rho(t) a_l^\dag] - e^{ikx_{jl}} [a_j^\dag, a_l \rho(t)] \bigg)
\end{equation}
where the last term describes the effective long-range interactions between the two resonators mediated by the waveguide. By separating the interaction term into coherent and incoherent parts, the effective master equation described in the main text is obtained.

\end{document}